\documentclass[11pt]{article}
\usepackage{graphicx}

\newcommand{\BABARPubYear}    {04}

\newcommand{\BABARConfNumber} {045}
\newcommand{\SLACPubNumber} {10633}
\newcommand{\LANLNumber} {0408087}

\input pubboard/babarsym

\long\def\inst#1{\par\nobreak\kern 4pt\nobreak
    {\it #1}\par\vskip 10pt plus 3pt minus 3pt}

\begin{document}
{\pagestyle{empty}

\begin{flushright}
\babar-CONF-\BABARPubYear/\BABARConfNumber \\
SLAC-PUB-\SLACPubNumber \\
hep-ex/\LANLNumber \\
August, 2004 \\
\end{flushright}

\par\vskip 5cm

% Title of the paper
\begin{center}
\Large \bf 
{\boldmath  Search for the  $D^*_{sJ}(2632)^+$ at BaBar }
\end{center}
\bigskip

\begin{center}
\large The \babar\ Collaboration\\
\mbox{ }\\
\today
\end{center}
\bigskip \bigskip

% Abstract
\begin{center}
\large \bf Abstract
\end{center}
We have performed a search for the $D^*_{sJ}(2632)^+$ state
recently reported by the SELEX
Collaboration at FNAL.
This preliminary
analysis makes use of an integrated luminosity of 125 ${\rm fb}^{-1}$ 
collected
by the \babar\  detector at the PEP-II asymmetric-energy $e^+e^-$ collider.
The resulting $\Ds \eta$ and $D^0 K^+$ mass spectra show no evidence
for the $D^*_{sJ}(2632)^+$ state. In addition, no signal is observed in the
$D^{*+} K_S$ mass spectrum.

\vfill
\begin{center}

Submitted to the 32$^{\rm nd}$ International Conference on High-Energy Physics, ICHEP 04,\\
16 August---22 August 2004, Beijing, China

\end{center}

\vspace{1.0cm}
\begin{center}
{\em Stanford Linear Accelerator Center, Stanford University, 
Stanford, CA 94309} \\ \vspace{0.1cm}\hrule\vspace{0.1cm}
Work supported in part by Department of Energy contract DE-AC03-76SF00515.
\end{center}

\newpage
} % end of pagestyle{empty}

% Input author list file
%
%author list removed temporarily to save trees 7/9/04 RNC
%
\begin{center}
\small

The \babar\ Collaboration,
\bigskip

%% author list as of 02-Jul-2004 (609 authors)
%
B.~Aubert,
R.~Barate,
D.~Boutigny,
F.~Couderc,
J.-M.~Gaillard,
A.~Hicheur,
Y.~Karyotakis,
J.~P.~Lees,
V.~Tisserand,
A.~Zghiche
\inst{Laboratoire de Physique des Particules, F-74941 Annecy-le-Vieux, France }
% ...

\end{center}\newpage

% The body of the paper starts here
\section{INTRODUCTION}
\label{sec:Introduction}
The SELEX Collaboration at FNAL
has recently reported the existence of a narrow state at a 
mass of 2632~\mevcc
decaying to $D^+_s \eta$~\cite{selex}. That analysis was based on a 
sample of about 500
$D^+_s$ events.
Evidence for the same state in the corresponding
$D^0 K^+$ mass spectrum was also presented. This work has generated
considerable 
theoretical interest~\cite{aa} because of the anomalous decay mode
and since the state appears to have a small width despite having a mass
significantly above $D^0 K$ threshold.

In the present analysis, inclusive production of the 
$D^+_s \eta$, $D^0 K^+$, and $D^{*+}K_S$ systems in $e^+e^-$ collisions near
10.58~\gev  center-of-mass energy is investigated in a search for the
$D^*_{sJ}(2632)^+$ state. All results are preliminary.

\section{DETECTOR AND DATASET}

This analysis is performed using a 125~${\rm fb}^{-1}$
data sample collected on or near the $\Upsilon(4S)$ resonance
with the \babar\  detector at the
\pep2 asymmetric-energy $e^+e^-$ storage rings.
The \babar\ detector,
a general-purpose, solenoidal, magnetic spectrometer, is
described in detail elsewhere \cite{babar}. 
Charged particles were detected
and their momenta measured by a combination of a drift chamber
and silicon vertex tracker, both operating within a
1.5-T solenoidal magnetic field. 
A ring-imaging Cherenkov detector is used for
charged-particle identification. Photons are detected and
measured with a CsI electromagnetic calorimeter.

\section{\boldmath $D^+_s \eta$ EVENT SELECTION}

A clean sample of $K^\pm$ candidates is obtained using particle
identification by requiring a Cherenkov photon yield and angle
consistent with the $K^\pm$ hypothesis. This information
is augmented with energy loss measurements in the tracking systems.
The efficiency of $K^\pm$ identification is approximately 85\% in the
kinematic range used in this analysis with a $\pi^\pm$ contamination
of less than 2\%. A similar procedure is used to produce a sample
of $\pi^\pm$ candidates.

Each $\Ds$ candidate\footnote{Inclusion of 
charge conjugate states is 
implied throughout this paper} is constructed by
combining a $\Kp \Km$ candidate pair with a $\pip$ candidate 
in a geometric
fit to a common vertex. An acceptable $\Kp\Km\pip$ candidate must have a fit
probability greater than 0.1\% and a trajectory consistent with 
originating from the $e^+e^-$ 
luminous region.
Backgrounds are further suppressed by selecting 
decays to $\Kbar^{*0}K^+$ and $\phi\pip$.
Additional details of this selection procedure
can be found elsewhere~\cite{Aubert:2003fg}.  

\begin{figure}[ht]
\vspace{-1cm}
\begin{center}
\includegraphics[width=10cm]{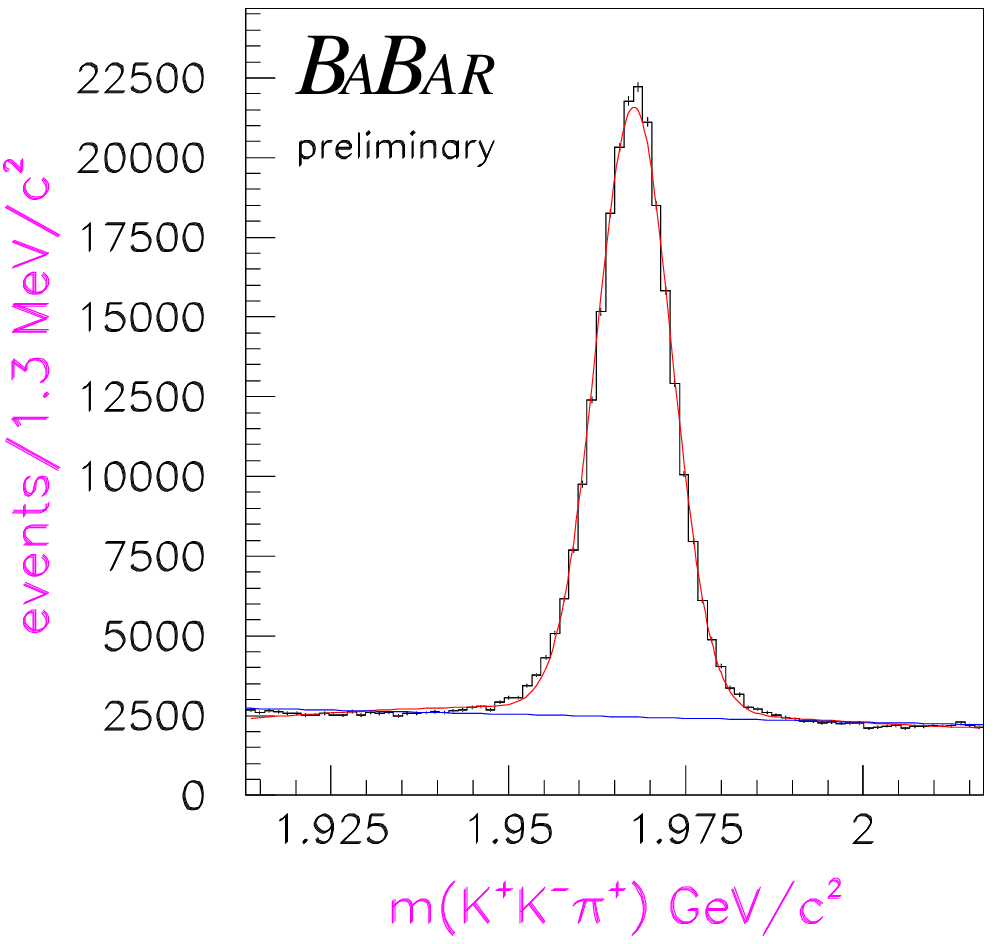}
\caption{\label{fig:kkpi} 
The $\Kp\Km\pip$ mass distribution after applying the selection procedure
described in the text.}
\end{center}
\end{figure}

The resulting $\Kp\Km\pip$ mass 
distribution is shown in Fig.~\ref{fig:kkpi}. 
The $D_s^+$ signal peak is centered at a mass of 1.968~\gevcc and has
rms deviation 5.2~\mevcc, as determined by a 
fit that includes a double-Gaussian representation of the signal with
a second-order polynomial to describe the background. 
The fit determines a yield of approximately
196,000 signal events.

For events containing a $\Ds$ candidate,
$\eta$ candidates are selected in the $\gamma \gamma$ decay mode.
It is assumed that each $\eta$ originates from the interaction point
({\rm i.e.}, the $D^*_{sJ}(2632)^+$ is short lived).
The $\eta$ signal-to-background ratio has been enhanced by means
of the following selection criteria: 
\begin{itemize}
\item{} Each $\gamma$ cannot be part of any $\pi^0$ which has
momentum greater than 150~\mevc.
\item{} Any $\gamma$ compatible with the decay $D^*_s(2112)^+ \to
D^+_s \gamma$ is removed.
\item{} Each $\gamma$ energy must be greater than 350~\mev, and 
the energy sum for a candidate $\gamma$ pair must be greater than 
1.15~\gev.
\item{} The quantity $|\cos\theta_\gamma|$ must be less than
$0.85$, where $\theta_\gamma$ is the
helicity angle of one $\gamma$ in the $\gamma\gamma$ rest frame
with respect to the $\eta$ candidate direction in the laboratory frame.
\end{itemize}

\begin{figure}[ht]
\vspace{-1cm}
\begin{center}
\includegraphics[width=10cm]{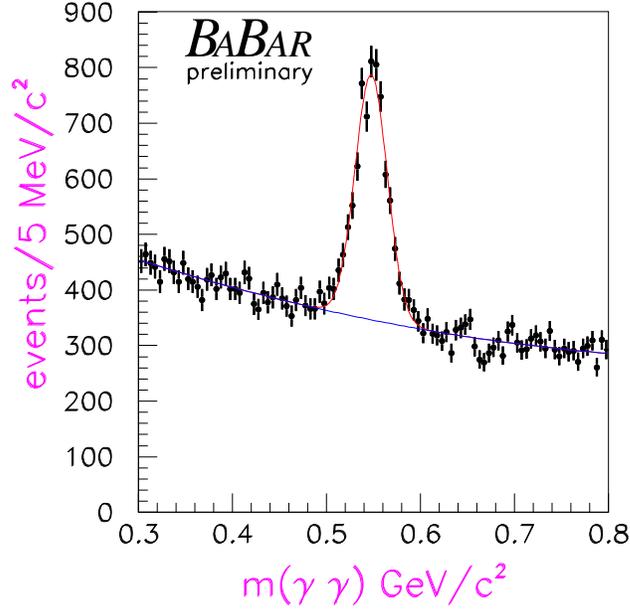}
\caption{\label{fig:eta} 
The $\gamma \gamma$ effective mass distribution in the $\eta$
region after the selection procedure described in the text. The presence
of a $\Ds$ candidate is required.} 
\end{center}
\end{figure}

The resulting $\gamma \gamma$ effective mass 
distribution is shown in Fig.~\ref{fig:eta}. A fit to the mass spectrum
using a Gaussian signal function and a second-order polynomial background
function yields the following parameter values (statistical
errors only) for the $\eta$:
\begin{equation}
m = [547.4 \pm 0.5]\;\mevcc \quad \sigma=[17.1 \pm 0.5]\;\mevcc \;.
\end{equation}
The mass value is in excellent agreement with the PDG value~\cite{PDG}. 
The resulting $\eta$ 
signal consists of approximately 3900 events.

\section{\boldmath THE $D^+_s \eta$ SYSTEM}

The $\gamma\gamma$ mass distribution of Fig.~\ref{fig:eta} is for events
containing a $\Ds$ candidate. However, the background under the
$\Ds$ signal (Fig.~\ref{fig:kkpi}) and the substantial background
under the $\eta$ signal (Fig.~\ref{fig:eta}) mean that it is not clear
whether there is any correlation between the $\Ds$ and $\eta$ signals.

\begin{figure}[ht]
\vspace{-1cm}
\begin{center}
\includegraphics[width=16cm]{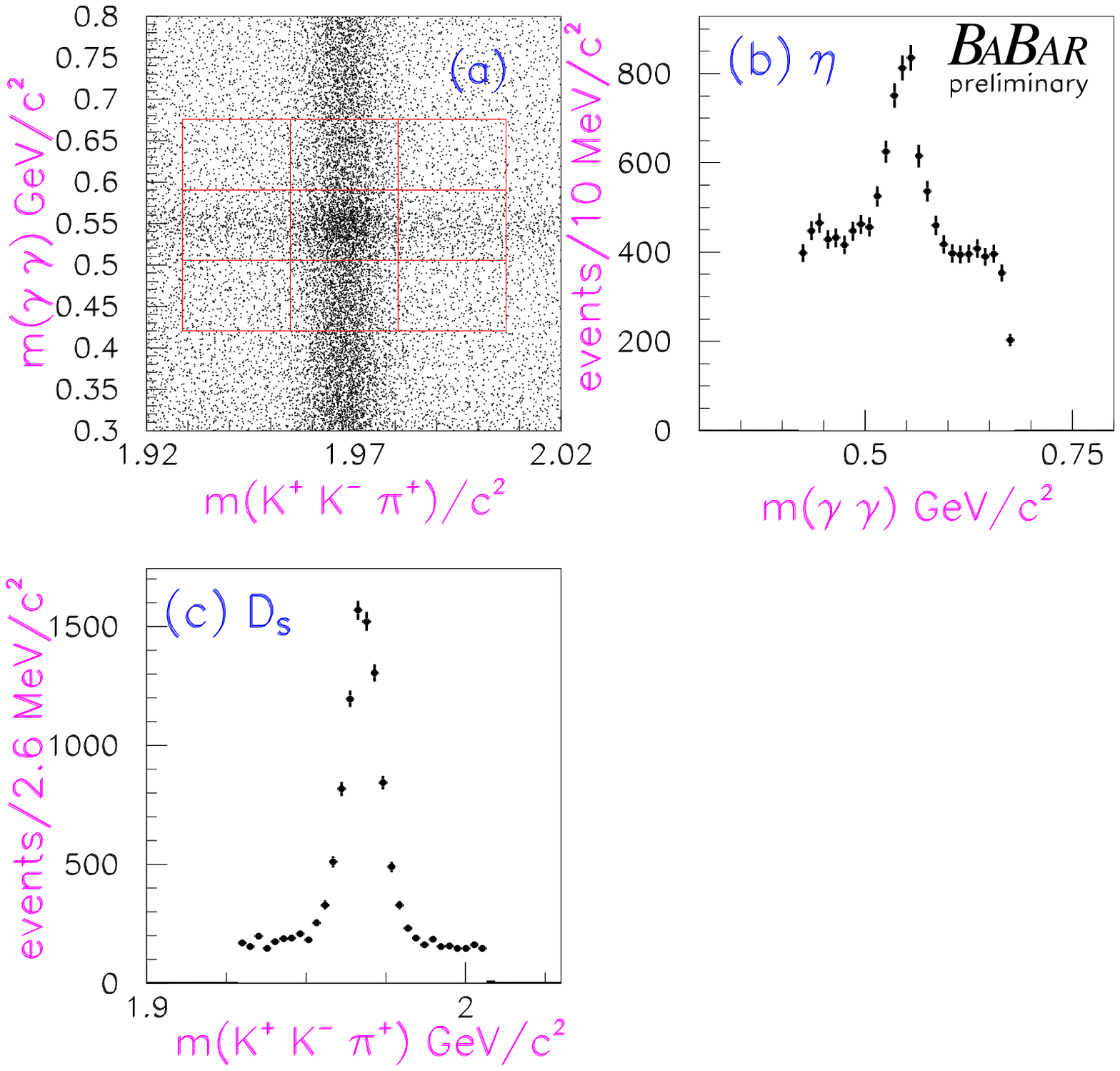}
\caption{\label{fig:scat} 
(a) The scatterplot of $m(\gamma \gamma)$ vs.
$m(K^+ K^- \pi^+)$ for $p^*(D_s^+ \eta)> 2.5$~\gevc. (b) The $\gamma \gamma$ 
and (c) $K^+ K^- \pi^+$ mass projections for the selected region.}
\end{center}
\end{figure}

Figure~\ref{fig:scat}a shows the scatterplot of $m(\gamma \gamma)$ versus
$m(K^+ K^- \pi^+)$ with the additional requirement that the 
$e^+e^-$ center-of-mass
momentum $p^*(D_s^+ \eta)$ of the $\Ds\eta$ system is at least 2.5~\gevc
to suppress background.
The $\eta$ and $\Ds$ signal regions are quite clear, and the distribution
appears rather uniform in the background regions and within the signal
bands, except in the region of overlap. In order to establish the presence
of an excess of events in the overlap region corresponding to correlated
$\Ds$ and $\eta$ production, we perform a two-dimensional subtraction.
The scatterplot is divided into the nine subregions of equal area indicated 
in Fig~\ref{fig:scat}a. These subregions are centered on the $\Ds$ and $\eta$
mass values and extend by plus or minus 2.5 standard deviations in each
mass variable.
Figure~\ref{fig:scat}b and Fig.~\ref{fig:scat}c 
show the $\gamma \gamma$ and 
the $K^+ K^- \pi^+$ mass projections, respectively, for the selected
mass region.

Labeling the subregions of Fig.~\ref{fig:scat}a from 1 to 9, from left
to right and bottom to top, the excess number of events $N(\Ds\eta)$ 
in the central
subregion (5) is estimated from the following linear equation:
\begin{equation}
N(\Ds\eta) = N_5 - \left( N_2 + N_4 + N_6 + N_8 \right)/2
+ \left(N_1+N_3+N_7+N_9\right)/4 \;,
\end{equation}
under the assumption (consistent with Fig.~\ref{fig:scat}a) that
any mass dependence in the selected region is at most linear.
This procedure yields the estimate
\begin{equation}
N(\Ds\eta) = 1102 \pm 75
\end{equation}
(statistical error only),
so that there is clear evidence of correlated $\Ds$ and $\eta$ production.

\begin{figure}[ht]
\vspace{-1cm}
\begin{center}
\includegraphics[width=16cm]{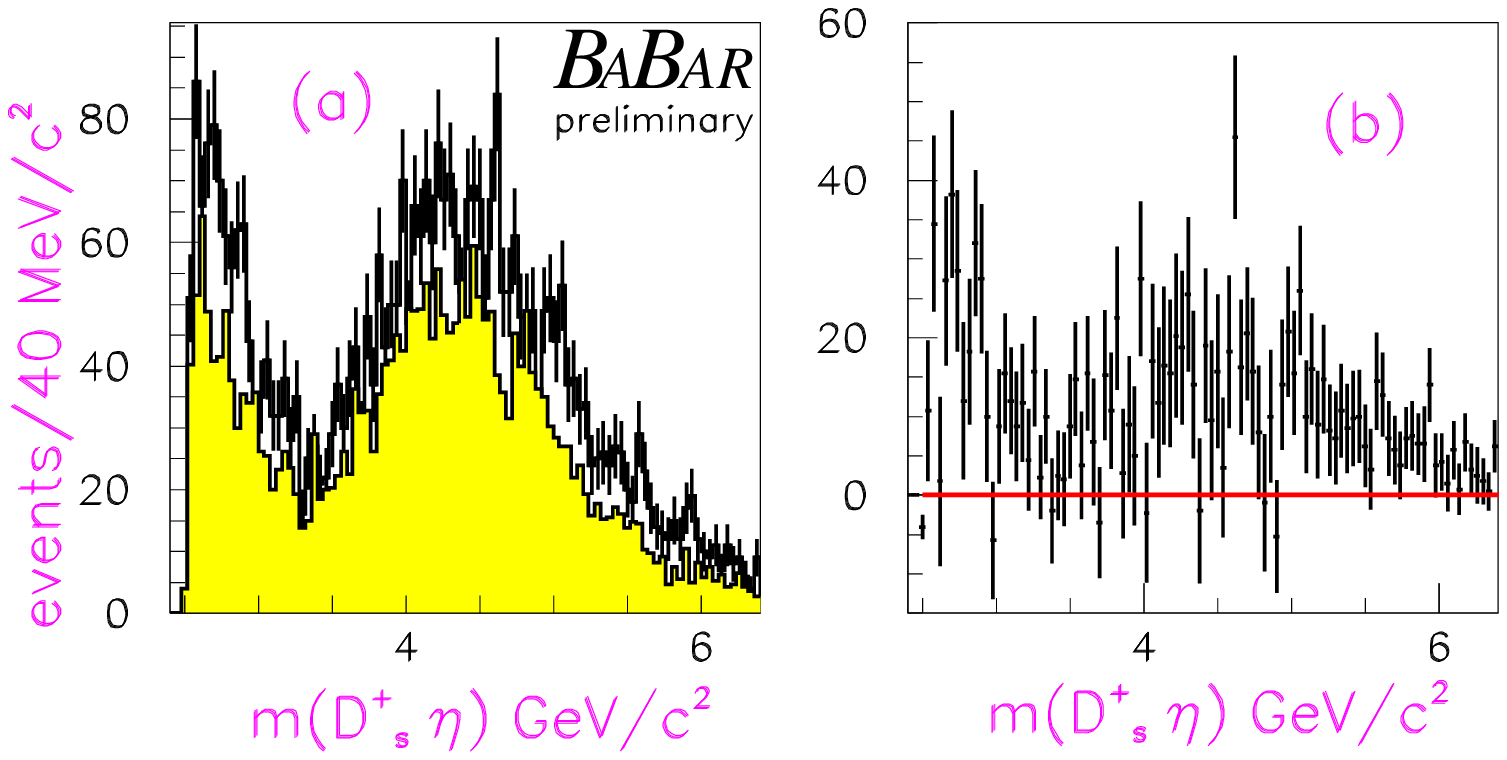}
\caption{\label{fig:dseta} 
(a) The $D^+_s \eta$ invariant mass distribution. The unshaded distribution
($m_5$) corresponds to the central region of Fig.~\ref{fig:scat}a  while the 
shaded
distribution is obtained using Eq.~\ref{eq:mshade}.
(b) The $D^+_s \eta$ mass distribution
obtained by subtracting the distributions of (a).} 
\end{center}
\end{figure}

In order to obtain the
$\Ds\eta$ mass distribution $m(\Ds\eta)$ corresponding to this excess,
a $\Ds\eta$ invariant mass distribution ($m_i$) is produced for each of the
nine subregions, $i$. In calculating the invariant mass, the candidate
$\Ds$ or $\eta$ three-momentum vector is combined with the relevant
PDG mass value to obtain the energy. The unshaded histogram of 
Fig.~\ref{fig:dseta}a shows this mass distribution for the center
subregion ({\it i.e.}, $m_5$), while the shaded histogram is obtained
from
\begin{equation}
m_b = 
\left( m_2 + m_4 + m_6 + m_8 \right)/2 - \left(m_1+m_3+m_7+m_9\right)/4 \;.
\label{eq:mshade}
\end{equation}
The distribution of Fig.~\ref{fig:dseta}b is obtained by subtracting
the shaded distribution of Fig.~\ref{fig:dseta}a
from the unshaded distribution. It follows that
this distribution corresponds to
correlated $\Ds\eta$ production under the assumption of linear signal
and background behavior.

There appear to be two distinct regions in this mass distribution.
The region above 3.5~\gevcc  can be interpreted as being the result of
continuum production of two (or more) jets with the $\Ds$ and $\eta$
produced in opposing jets. The region below 3~\gevcc shows a
monotonic rise toward threshold. This is interpreted as being the
result of $\Ds$ and $\eta$ production within a single jet, and hence
is the region in which any resonant structures in $\Ds\eta$ mass
should be seen.

\begin{figure}[ht]
\vspace{-1cm}
\begin{center}
\includegraphics[width=12cm]{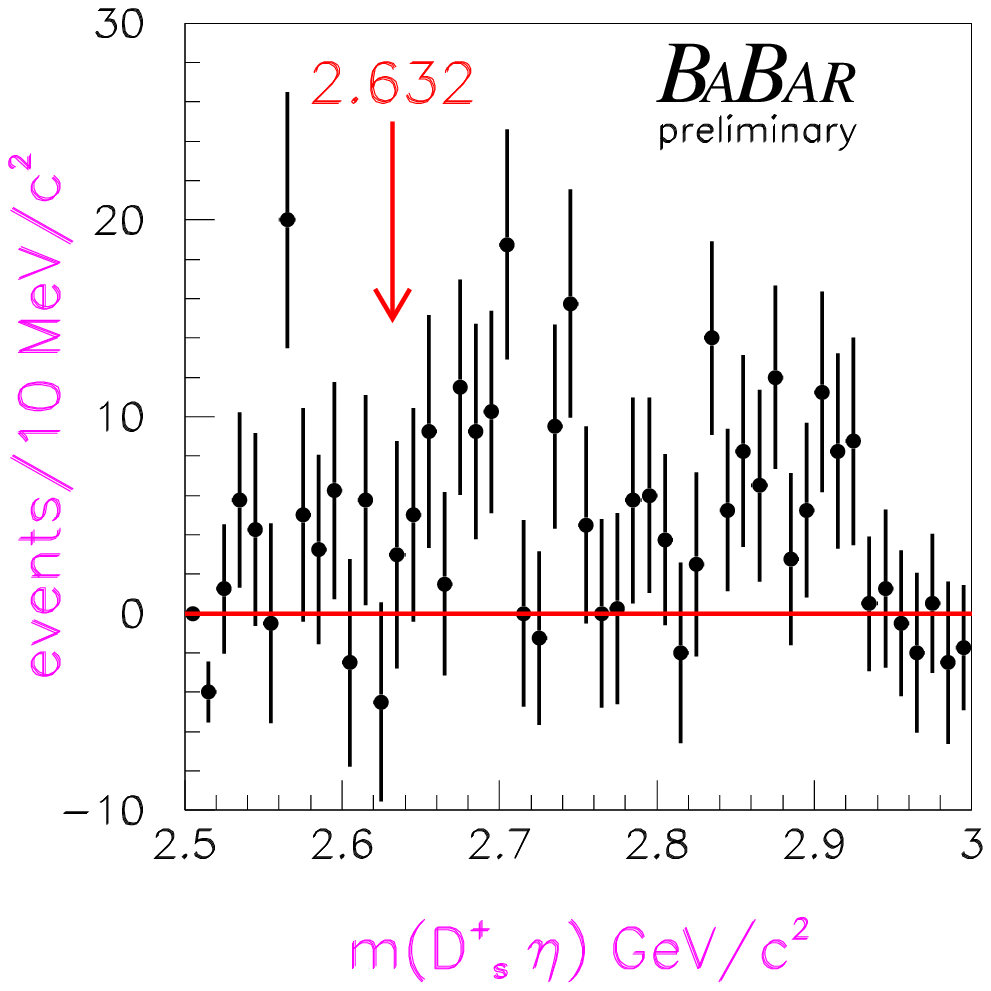}
\vspace*{8pt}
\caption{\label{fig:dseta_small} 
The $D^+_s \eta$ invariant mass distribution of Fig.~\ref{fig:dseta}b
for the region below 3~\gevcc. The arrow indicates the mass location at which
the $D^*_{sJ}(2632)^+$ state should appear.}
\end{center}
\end{figure}

The mass region of Fig.~\ref{fig:dseta}b below 3.0~\gevcc  is shown
in detail in Fig.~\ref{fig:dseta_small}. The arrow indicates the
location at which the $D^*_{sJ}(2632)^+$ state should appear. 
There is no evidence for a signal.

\begin{figure}[ht]
\vspace{-1cm}
\begin{center}
\includegraphics[width=12cm]{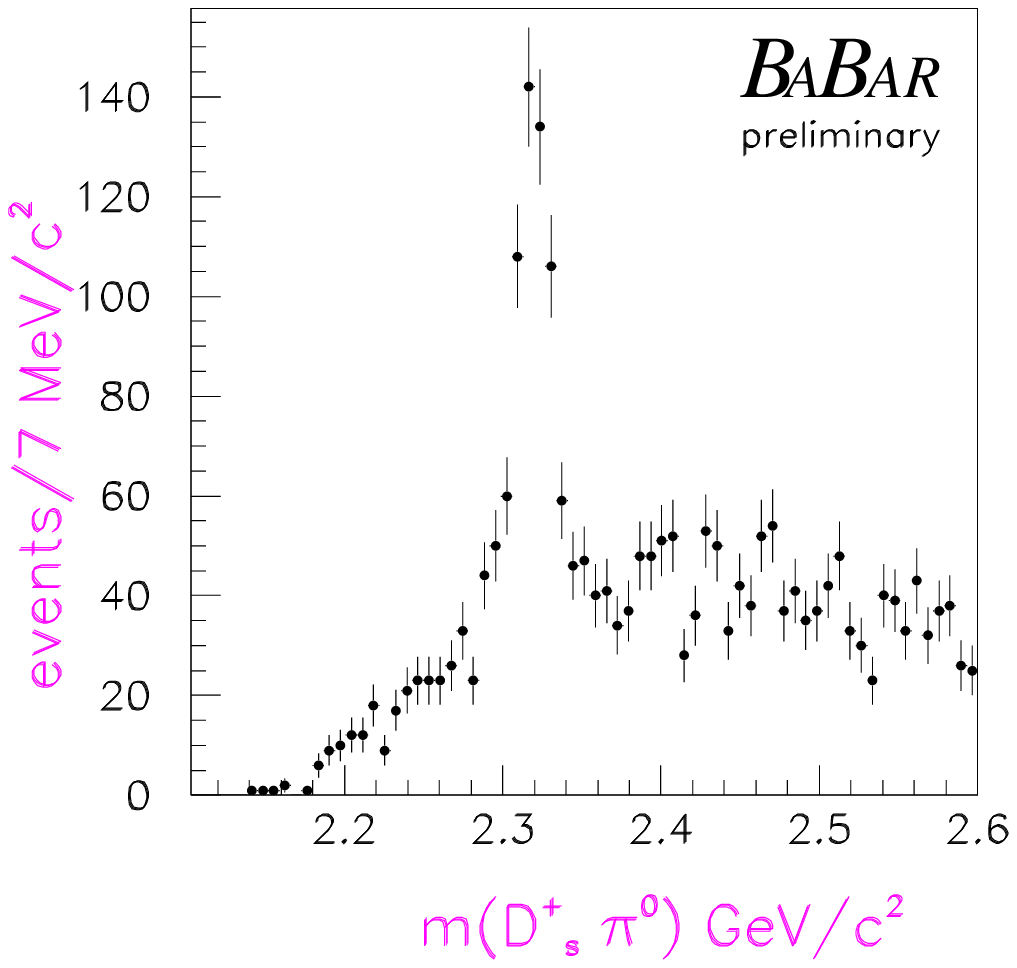}
\vspace*{8pt}
\caption{\label{fig:dspi0} 
The $D^+_s \pi^0$ invariant mass spectrum for selection
criteria similar to those used in the $\Ds\eta$ candidate selection.}
\end{center}
\end{figure}

The requirements imposed on the selection of the $\eta$ candidates are 
rather stringent, but are not expected to entirely remove any 
$D^*_{sJ}(2632)^+$ signal.
As a check, we use similar requirements to select $\Ds\piz$ candidates.
The resulting $\Ds\piz$ mass spectrum is shown in Fig.~\ref{fig:dspi0}.
A large $D_{sJ}^*(2317)^+$ signal is observed.

\section{\boldmath THE $D^0 K^+$ SYSTEM}

The $D^0 K^+$ mass spectrum has been investigated using the 
$D^0 \to K^- \pi^+$ decay mode. 

A $D^0$ candidate is constructed by combining a
$\pi^+$-$K^-$ pair in a geometric fit to a common vertex. An acceptable
candidate must have a fit
probability greater than 1\% and a trajectory consistent with 
originating from the $e^+e^-$ luminous region. In addition, the
$D^0$ candidate must have $p^* > 0.5$~\gevc.

\begin{figure}[ht]
\vspace{-1cm}
\begin{center}
\includegraphics[width=10cm]{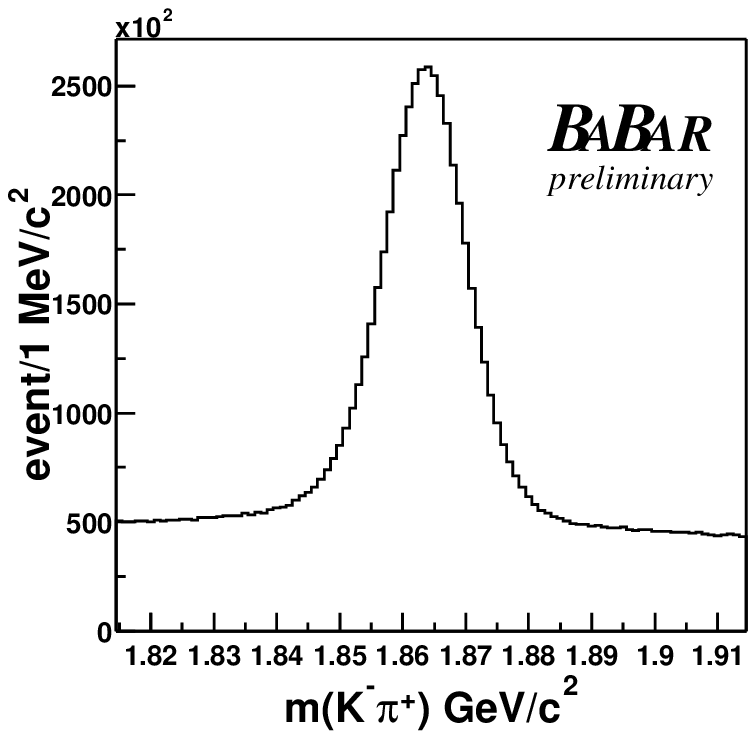}
\vspace*{8pt}
\caption{\label{fig:d0} 
The $K^- \pi^+$ mass distribution after applying 
the selection procedure described in the text.}
\end{center}
\end{figure}

The resulting $\pip\Km$ mass distribution for the $D^0$ mass region
is shown in Fig.~\ref{fig:d0}. The signal peak is centered at 
1.864~\gevcc and has an rms deviation of 8.2~\mevcc. There are approximately
$3.7 \times 10^6$ signal events above background.

A $D^0$ candidate with mass within $20$~\mevcc of the central value
is combined with a well-identified $\Kp$ track in a fit to a common
vertex. The vertex fit probability must be greater than 1\% and the
vertex position must be consistent with the $e^+e^-$ luminous region.

\begin{figure}[ht]
\begin{center}
\includegraphics[width=10cm]{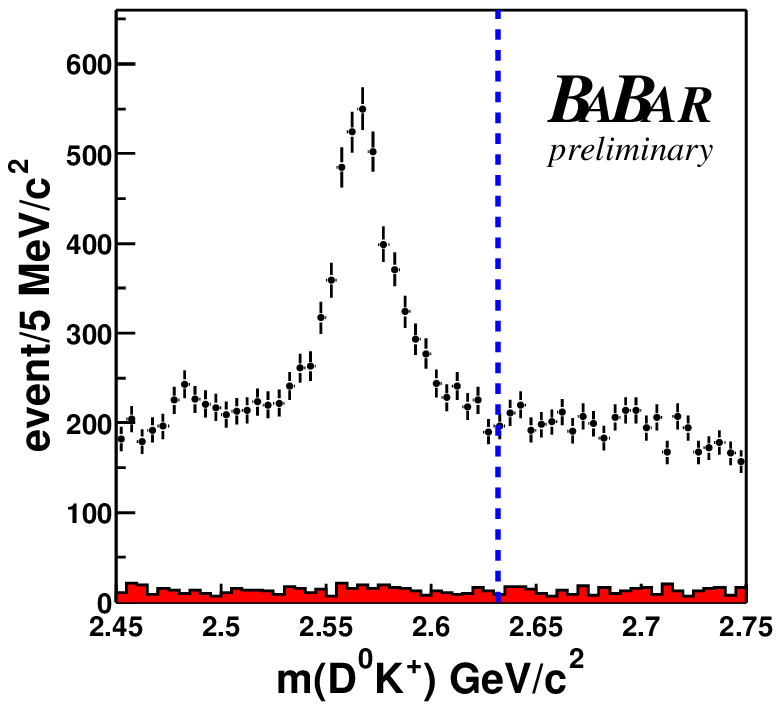}
\vspace*{8pt}
\caption{\label{fig:d0k} 
The $D^0 K^+$ invariant mass distribution after applying the 
selection procedure
described in the text. The dashed line indicates the location at which
the $D^*_{sJ}(2632)^+$ state should appear.
The shaded histogram is the mass distribution for 
wrong-sign $D^0 K^-$ pairs.}
\end{center}
\end{figure}

Requiring $p^*(D^0 K^+)>4.0$~\gevc we obtain the $D^0\Kp$ mass spectrum shown 
in Fig.~\ref{fig:d0k}. The large peak is due to the 
decay $D_{s2}(2573)^+ \to D^0 K^+$; the skewing of the signal toward
high mass is consistent
with a spin 2 interpretation of this state.
The shaded histogram is the mass distribution for 
wrong-sign $D^0 K^-$ pairs.
There is no evidence for structure in the 2.632~\gevcc mass region.

\section{\boldmath THE $D^{*+} K_S$ SYSTEM}

\begin{figure}[ht]
\vspace{-1cm}
\begin{center}
\includegraphics[width=10cm]{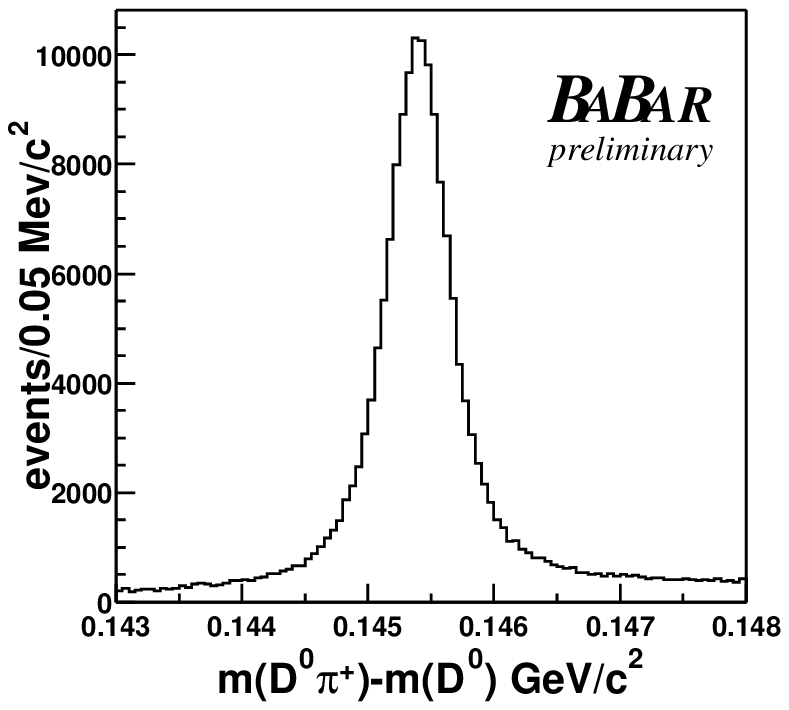}
\vspace*{8pt}
\caption{\label{fig:dm_DstarKs} 
The $\delta M$ mass distribution after applying 
the selection procedure described in the text.}
\end{center}
\end{figure}

A $D^0$ candidate with mass within 25~\mevcc of the central mass value
(Fig.~\ref{fig:d0}) is combined with a well-identified $\pip$ track
in a fit to a common vertex. The vertex fit probability must be greater
than 1\% and the vertex position must be consistent with
the $e^+e^-$ luminous region. We define the difference
$\delta M$ between the $\Dz\pip$ and $\Dz$ invariant mass values by:
\begin{equation}
\delta M = \sqrt{ \left( p_{K^-} + p_{\pi_1^+} + p_{\pi_2^+} \right)^2
           - \left( p_{K^-} + p_{\pi_1^+} \right)^2 } \;,
\end{equation}
where $\pi_1^+$ is from the $\Dz$ candidate and $\pi_2^+$ is from the
$D^{*+}$ candidate. The distribution of $\delta M$ for the
$D^*(2010)^+$ region is shown in
Fig.~\ref{fig:dm_DstarKs}; there is a clear $D^{*+}$ signal consisting
of approximately $1.4\times 10^5$ events over a small background. We require
a $D^{*+}$ candidate to have $\delta M$ in the interval
$145.4\pm 1.6$~\mevcc.

\begin{figure}[ht]
\begin{center}
\includegraphics[width=10cm]{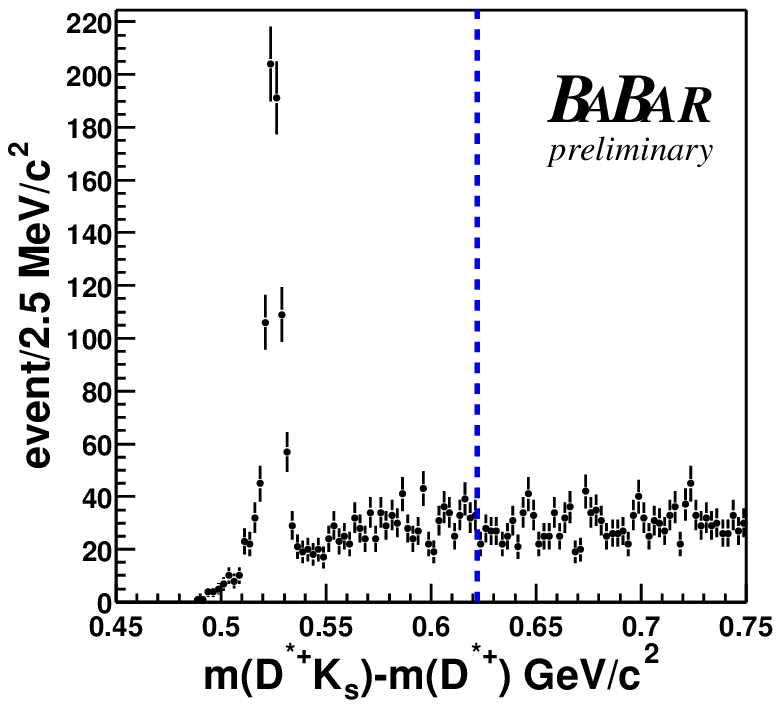}
\vspace*{8pt}
\caption{\label{fig:dstarks} 
The $D^{*+} K_S$ invariant mass distribution after applying the 
selection procedure
described in the text. The dashed line indicates the mass location 
at which the $D^*_{sJ}(2632)^+$ state might appear.}
\end{center}
\end{figure}

A candidate $K_S$ track is reconstructed by vertexing a well-identified
$\pip\pim$ pair. The vertex fit probability is required to be greater
than 1\%, and the $\pip\pim$ invariant mass must be within 16~\mevcc
of the $K_S$ PDG mass value~\cite{PDG}. The candidate $K_S$ trajectory
is then required to be consistent with the vertex of a $D^{*+}$ candidate
such that the $K_S$ flight length exceeds 1~mm. The distribution of the 
resulting
difference in invariant mass between the $D^{*+}K_S$ and $D^{*+}$
track combinations is shown in Fig.~\ref{fig:dstarks} with the requirement
that $p^*(D^{*+}K_S) > 4$~\gevc. The large, narrow peak just above threshold
results from production of the $D_{s1}(2536)^+$. The vertical dashed line
indicates the mass position at which the $D^*_{sJ}(2632)^+$ state 
might be observed. There is no evidence for production of this state.

\section{SUMMARY}

The SELEX Collaboration has reported the existence of a charm meson state,
the $D^*_{sJ}(2632)^+$, with $D^+_s \eta$ and $D^0K^+$ decay modes.
We have searched for this state using $e^+e^- \to c\overline{c}$ collision
data from 125~${\rm fb}^{-1}$ of integrated
luminosity collected by the \babar\  experiment.
In this preliminary analysis 
we find no evidence for this state in inclusive production of
$D^+_s \eta$, $D^0K^+$, or $D^{+*}K_S$.

\section*{ACKNOWLEDGMENTS}
\label{sec:Acknowledgments}

% Specific acknowledgments for this paper; remove if not needed.

% Standard acknowledgments paragraph; must always be included.
We are grateful for the 
extraordinary contributions of our \pep2\ colleagues in
achieving the excellent luminosity and machine conditions
that have made this work possible.
The success of this project also relies critically on the 
expertise and dedication of the computing organizations that 
support \babar.
The collaborating institutions wish to thank 
SLAC for its support and the kind hospitality extended to them. 
This work is supported by the
US Department of Energy
and National Science Foundation, the
Natural Sciences and Engineering Research Council (Canada),
Institute of High Energy Physics (China), the
Commissariat \`a l'Energie Atomique and
Institut National de Physique Nucl\'eaire et de Physique des Particules
(France), the
Bundesministerium f\"ur Bildung und Forschung and
Deutsche Forschungsgemeinschaft
(Germany), the
Istituto Nazionale di Fisica Nucleare (Italy),
the Foundation for Fundamental Research on Matter (The Netherlands),
the Research Council of Norway, the
Ministry of Science and Technology of the Russian Federation, and the
Particle Physics and Astronomy Research Council (United Kingdom). 
Individuals have received support from 
CONACyT (Mexico),
the A. P. Sloan Foundation, 
the Research Corporation,
and the Alexander von Humboldt Foundation.

\end{document}